\documentclass[11pt,a4paper]{article}
\usepackage{amsfonts,amsmath,amstext,amssymb}
\usepackage{dsfont}
\usepackage{theorem}    
\usepackage{a4wide}

\usepackage{graphicx}

\DeclareGraphicsExtensions{.bmp,.png,.pdf,.jpg,.JPEG}

\let\ddpp\mathds 
\def\R{\ddpp R}     
\def\S{\ddpp S}     
\def\H{\ddpp H}     
\def\L{\ddpp L}     

\newtheorem{lema}{Lemma}[section]

\newtheorem{teor}[lema]{\bf Theorem}

\newtheorem{rema}[lema]{\bf Remark}

\theorembodyfont{\rmfamily}

\title{An analytical approach to the external force-free motion of pendulums on surfaces of constant curvature}

\author{
Rafael M. Rubio${}^{a*}$ and  Juan J. Salamanca${}^{a*}$\\[6mm]
${}^a$ Departamento de Matem\'aticas, Campus de Rabanales, \\[0.5mm]
Universidad de C\'ordoba, 14071 C\'ordoba, Spain,\\[0.5mm]
E-mails\textup{:\texttt{\;rmrubio@uco.es},\,\,\texttt{jjsalamanca@uco.es}}}

\date{}

\begin{document}

\maketitle

\begin{abstract}
The dynamics of force free  motion of pendulums on  surfaces of constant Gaussian curvature is addressed when the pivot moves along a geodesic obtaining 
the Lagragian of the system. As a application it is possible the study of elastic and quantum pendulums.
\end{abstract}

\vspace{2mm} 

\noindent{\it AMS classification scheme numbers}: {70H03, 53A35, 53B20.}

\vspace{1pc}

\maketitle

\section{Introduction}

The classical theory of force-free motions of rigid particle systems has a long history in connection with investigation
in Mathematics and Mechanics. From the Differential Calculus, in the case of translational particle motions in
Euclidean space (Newton axioms of Mechanics), to the Riemann Geometry. Hence, the definition of geodesics has immediately concerned
with force-free particle motions on surfaces. The introduction of Riemannian manifolds and the geometry of their geodesics were
motivated by the mechanics of constrained particle systems. In the last few years, the study of force-free motion systems on 
Riemannian manifolds of constant sectional curvature has attracted the interest of several authors \cite{carinera1}, \cite{C-F-G}, \cite{C-F-G2}, \cite{M}, \cite{N} and
\cite{S}.

In particular, in \cite{C-F-G} the authors define the notion of a pendulum on a surface of constant Gaussian curvature $K$ and they study the motion of a mass
at a fixed distance from a pivot. So, a pendulum problem on a surface of constant curvature is defined as a pivot point and a mass connected to
that point
by a rigid massless rod of fixed length $\rho$. It is assumed that the pivot is constrained to move along some fixed curve with prescribed motion.
The rod provides the only force on the mass in order to keep
the mass at the fixed distance from the pivot.
No torque is applied to the rod, (fig. 1). 
In \cite{C-F-G} it is studied the pendulum problem when the pivot moves along a geodesic path, and the
space is a surface of constant (not zero) curvature. It is considered the surface immersed in $\R^3$ or $\L^3$ and it is obtained the
differential motion equation by Newtonian procedure doing a laborious calculation. Moreover, the cases $K>0$ and $K<0$ are necessary to come
from different forms.

In this work, we deal with the pendulum problem from an analytic point of view.
 This procedure allows approach simultaneously the cases of positive
and negative curvature and also of zero curvature (as a limit case), with very simple computations. 
The key of our study is a lagragian approach,  which has a 1-dimensional configuration space.
As adirect consequence the internal curvature force is conservative and its potential function is easily calculated.
 Moreover, this method can be used
to another related problems, as the elastic pendulum, or the quantum pendulum.

Mainly, this work concerns the case in which the pivot moves along a geodesic. Let $\zeta$ be the angle between the rod
 and the motion direction of the pivot. We will assume the following convention:

\vspace{1mm}

\noindent If the constant curvature of the space is $K>0$,

$$\cos_K(z)=\cos(z\sqrt{K}),\quad \sin_K(z)=\sin(z\sqrt{K}),\quad \tan_K(z)=\tan(z\sqrt{K}) \, .$$

\noindent If the constant curvature of the space is $K<0$,

$$\cos_K(z)=\cosh(z\sqrt{-K}),\quad \sin_K(z)=\sinh(z\sqrt{-K}),\quad \tan_K(z)=\tanh(z\sqrt{-K})$$

The main result is the following:

{\quote {\it Suppose that the pivot of pendulum moves with constant speed $v$ along a geodesic on a surface with constant curvature $K$.
 Let $\zeta(t)$ the angle at the time $t$ between the rigid rod and the direction of pivot motion. Then the Lagrangian of the system is
$${\cal L}(\zeta,\dot\zeta)=\frac{1}{2}m\dot{\zeta}^2\sin_K^2(\rho)\frac{1}{\mid K\mid}-\frac{1}{2}mv^2{\rm sgn}(K)\sin_K^2(\rho)\sin^2(\zeta).$$

\noindent Therefore, the motion equation is

\begin{equation} \label{P}
\frac{d^2\zeta}{dt^2}=-v^2K\sin_K(\zeta)\cos_K(\zeta).
\end{equation}}}

Using the previous result and geometric arguments, we obtain the motion equation of the system when the pivot accelerates
along a geodesic (see Section 4). Further developements are discussed in Section 5.

\begin{figure}
\centering
\begin{tiny}

\end{tiny}\includegraphics{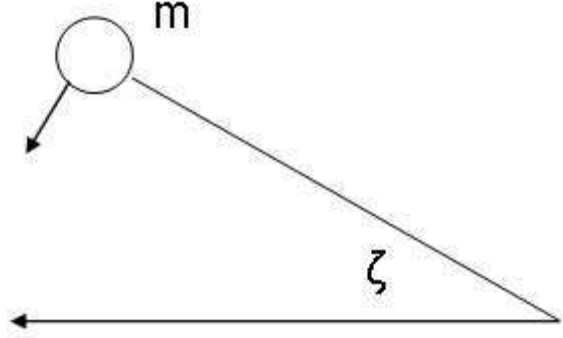}
\caption{Pendulum motion}
\label{fig 1}
\end{figure}

\section{Preliminaries} 

It is well known that a complete, simply-connected Riemannian $n$-manifold with constant sectional curvature
is isometric to one of model spaces $\R^n$, $\S^n(R)$ or $\H^n(R)$ (see \cite{L}).

\vspace{2mm}

 Let $(\S^2 (R),\langle\,,\,\rangle_1)$ be the 2-dimensional sphere  of radius $R$,
 endowed with the metric induced from $(\R^3, \langle\,,\,\rangle)$, where
$$\langle\,,\,\rangle=dx^2+dy^2+dz^2.$$

\noindent Consider the coordinate system $\{U_1, \Phi=(\varphi,\theta)\}$ in $S^2(R)$ with $U_1=\Phi^ {-1}((0,\pi)\times (0,2\pi))$ and
$\Phi^{-1}(\varphi,\theta)=(x,y,z)$, being

$$\left.
\begin{array}{l}
  x=R\sin{\varphi}\cos{\theta}
  \\ y=R\sin\varphi\sin\theta
\\ z=R\cos\varphi
\end{array}
\right\}$$

\noindent In this coordinate system the metric is given by
$$\langle\,,\,\rangle_1=R^2\, d\varphi\otimes d\varphi + R^2\sin^2\varphi\, d\theta\otimes d\theta,$$

\noindent and the non-zero Christoffel symbols are $\Gamma_{\varphi\theta}^{\theta}=\cot\varphi$ and
$\Gamma_{\theta\theta}^{\varphi}=-\sin\varphi\cos\varphi$.

\vspace{2mm}

Let's consider also the hyperbolic plane $(\H^2(R),\langle\,,\,\rangle_2)$, where $$\H^2(R)=\{(x,y,z)\in\R^3/x^2+y^2-z^2=-R^2\}$$

\noindent and $\langle\,,\,\rangle_2$ is the induced metric from the Lorentz-Minkowski spacetime $\L^3$.

\noindent Let $\{U_2, \Psi=(\alpha,\beta)\}$ be the coordinate system  in $\H^2(R)$ with $U_2=\Psi^ {-1}((-\infty,\infty)\times(-\infty,\infty))$ and
$\Psi^{-1}(\alpha,\beta)=(x,y,z)$, where

$$\left.
\begin{array}{l}
  x=R\cosh{\alpha}\sinh{\beta}
  \\ y=R\sinh\alpha
\\z=R\cosh\alpha\cosh\beta
\end{array}
\right\}$$

\noindent The metric is given by

$$\langle\,,\,\rangle_2=R^2d\alpha\otimes d \alpha+R^2\sinh^2 \alpha \, d\beta\otimes d\beta$$

\noindent and the non-zero Christoffel symbols are
$\Gamma_{\alpha\beta}^{\beta}=\coth\alpha$ and $\Gamma_{\beta\beta}^{\alpha}=-\sinh\alpha\cosh\beta$.

\section{Rigid pendulum}
Firstly, we deal with the rigid pendulum problem in the case that the pivot moves along a geodesic of its space, with constant speed.
 We will do analogously the study for
$K>0$ and $K<0$.

\subsection{ Spherical case, kinetic energy}

Note that in the case $K>0$ we must require that $\rho<\pi R$ to guarantee that the mass and the rod are on the same side of the geodesic line.

Suppose that the pivot moves along the geodesic $\varphi=\frac{\pi}{2}$ with constant speed. Let's take a reference associated to the pivot. Denote by 
$\zeta$ the angle between the rigid rod and the direction of pivot motion.

The Lagrangian of the system has two components, on the one hand, the kinetic energy of the mass ${\bf T}$, and on the other hand,
the potential energy ${\bf V}$, which is due to the curvature of space, i.e., ${\bf V}=0$ if the space is flat.

Consider the isosceles geodesic triangle formed in an infinitesimal temporal interval, by the rod and the arc $dl$ traced by the mass (fig. 2).

\begin{figure}
\centering
\begin{tiny}

\end{tiny}\includegraphics[width=0.47\linewidth]{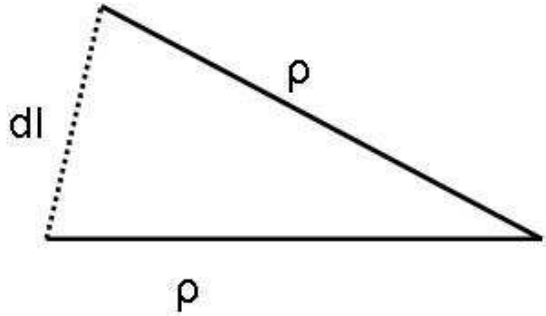}
\caption{Infinitesimal triangle}
\label{fig 2}
\end{figure}

 The Sine Theorem
for Spherical Geometry allows us to write
$$dl=R\,\sin(\rho/R)\,d\zeta=\frac{1}{\sqrt{\mid K\mid}}\sin(\rho\sqrt{\mid K\mid})\,d\zeta,$$

\noindent thus
$$\tilde v=\frac{dl}{dt}=\dot\zeta\sin(\rho\sqrt{\mid K\mid})\frac{1}{\sqrt{\mid K\mid}}.$$

\noindent Therefore, if the particle has mass $m$, the kinetic energy is given by

$${\bf T}=\frac{1}{2}m\dot\zeta^2\sin^2(\rho\sqrt{\mid K\mid})\frac{1}{{\mid K\mid}}.$$

\subsection{Spherical case, potential energy}

Consider the coordinate system $\{U_1, \Phi=(\varphi,\theta)\}$ with the metric $\langle\,,\,\rangle_1=R^2\, d\varphi\otimes d\varphi + R^2\sin^2\varphi\, d\theta\otimes d\theta,$ and suppose that the pivot moves
along the geodesic $\varphi=\frac{\pi}{2}$ with constant speed $v$. Obtain us the mass acceleration due to the curvature, so
 suppose that the mass moves along the curve

$$\left.
\begin{array}{l}
  \varphi=\varphi_0
  \\ \theta=\frac{v}{R}t
\end{array}
\right\}$$

\noindent Its acceleration is given by
$$a=\frac{v^2}{R^2}\Gamma_{\theta\theta}^\varphi \frac{\partial}{\partial\varphi}=-\frac{v^2}{R^2}\sin(\varphi)\cos(\varphi)\frac{\partial}{\partial\varphi}.$$

As consequence, the external force that we must be to apply on the rod will be ${\bf F}_e=ma$ and the curvature potential energy ${\bf V}$ satisfies ${\bf F}=-{\rm grad}\,({\bf V})$, 
where ${\bf F}=-{\bf F}_e$. Hence,

$${\bf V}=\int_{\pi/2}^{\varphi} R^2(-m\sin(\varphi)\cos(\varphi))\frac{v^2}{R^2}\,d\varphi=\frac{m}{2}v^2\cos^2\varphi$$

\noindent where we have taken $\pi/2$ as origin energy. Again, using the Sine Theorem (fig. 3) we obtain

\begin{figure}
\centering
\includegraphics[width=0.47\linewidth]{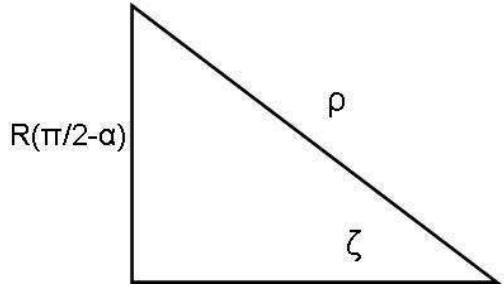}
\caption{Geodesic triangle}
\label{fig 3}
\end{figure}

$$\cos (\varphi)=\sin(\zeta)\sin(\rho/R) .$$

Thus

$${ \bf V}(\zeta)=-\frac{m}{2}v^2\sin^2(\rho/R)\sin^2(\zeta).$$

Now, we can enunciate the following Theorem,

\begin{teor} Suppose that the pivot of pendulum moves with constant speed $v$ along a geodesic on a surface with constant curvature $K>0$.
 Let $\zeta(t)$ the angle at the time $t$ between the rigid rod and the direction of pivot motion. Then the Lagrangian of the system is

\begin{equation}
{\cal L}(\zeta,\dot\zeta)=\frac{1}{2}m\dot{\zeta}^2\sin^2(\rho/R)\frac{1}{K}-\frac{1}{2}mv^2\sin^2(\rho/R)\sin^2(\zeta)
\end{equation}
\noindent Therefore, the motion differential equation is given by

\begin{equation}\label{motion1}
\frac{d^2\zeta}{dt^2}=-v^2K\sin(\zeta)\cos(\zeta)
\end{equation}

\end{teor}

If we do $K\rightarrow 0$,
$$\sin(\rho)\simeq \rho\sqrt{\mid K\mid}$$

\noindent and

$$\lim_{K\rightarrow 0}{\cal L}=\frac{1}{2}m \dot\zeta^2\rho^2,$$

\noindent obtaining the Lagrangian function of the pendulum moving in the Euclidean plane, when the pivot moves along a straigh line with constant speed.

\subsection{ Hyperbolic case, kinetic energy}

Consider us now,  the pivot moving along the geodesic $\alpha=0$, with constant speed $v$. Analogously to the spherical case, we take a reference
joint with the pivot and we denote $\zeta$ the angle between the rigid rod and the direction of pivot motion.

If we consider the isosceles hyperbolic differential triangle (fig. 2), and making use of the corresponding theorems of the Hyperbolic Geometry,
 we can to conclude that the kinetic energy of the system is given by

$${\bf T}=\frac{m}{2}\dot\zeta^2\sinh (\rho/R)\,\frac{1}{\mid K\mid}.$$

\subsection{ Hyperbolic case, potential energy}

Reasoning as the in spherical case, we consider that the pivot moves along the geodesic $\alpha=0$. If the mass moves along the curve

$$\left.
\begin{array}{l}
  \alpha=\alpha_0
  \\ \beta=\frac{v}{R}t
\end{array}
\right\},$$

\noindent we can to calculate the curvature potential function,

$${\bf V}=\int_{0}^\alpha R^2(-m\sinh(\alpha)\cosh(\alpha))\frac{v^2}{RÂ²}=-\frac{m}{2}v^2\sinh(\alpha)  \, ,$$

\noindent where we have taken $0$ as origin energy. Again, using the Sine Theorem in the hyperbolic case, we obtain
$$\sinh(\alpha)=\sin(\zeta)\sinh(\rho/R).$$

Thus,

$${\bf V}=-\frac{m}{2}v^2\sinh^2(\rho/R)\sin^ 2(\zeta) .$$

\begin{teor} Suppose that the pivot of the pendulum moves with constant speed $v$ along a geodesic on a surface with constant curvature $K<0$.
 Let $\zeta(t)$ the angle at the time $t$ between the rigid rod and the direction of pivot motion. Then the Lagrangian of the system is given by

\begin{equation}
{\cal L}(\zeta,\dot\zeta)=\frac{1}{2}m\dot{\zeta}^2\sinh^2(\rho/R)\frac{1}{\mid K\mid}+\frac{1}{2}mv^2\sinh^2(\rho/R)\sin^2(\zeta) .
\end{equation}
\noindent Therefore, the motion differential equation is

\begin{equation}\label{motion}
\frac{d^2\zeta}{dt^2}=v^2\mid K\mid\sinh(\zeta)\cosh(\zeta) .
\end{equation}

\end{teor}

With our convention, we can to enunciate (compare with \cite{C-F-G}, Theorem A):

\begin{teor}

Suppose that the pivot of pendulum moves with constant speed $v$ along a geodesic on a surface with constant curvature $K$.
 Let $\zeta(t)$ the angle at the time $t$ between the rigid rod and the direction of pivot motion. Then the Lagrangian of the system is given by
$${\cal L}(\zeta,\dot\zeta)=\frac{1}{2}m\dot{\zeta}^2\sin_K^2(\rho)\frac{1}{\mid K\mid}-\frac{1}{2}mv^2{\rm sgn}(K)\sin_K^2(\rho)\sin^2(\zeta).$$

\noindent Therefore, the motion differential equation is

$$\frac{d^2\zeta}{dt^2}=-v^2K\sin_K(\zeta)\cos_K(\zeta).$$

\end{teor}

\begin{rema} {\rm
Suppose $K>0$ and consider the motion equation (\ref{motion1}). Making the change $u=2\zeta$, we obtain the differential equation

\begin{equation}\label{planar}
\frac{d^2u}{dt^2}=-v^2 K\sin(u) ,
\end{equation}

\noindent which is the equation of planar pendulum of length $1$ in Euclidean space subject to a constant gravitational field of magnitude $g=v^ 2 K$. Its stable and unstable equilibria at $u=0$ and $u=\pi$, 
correspond to the stable $\zeta=0$ and unstable $\zeta=\pi/2$ equilibria of the pendulum on the spherical surface.

On the other hand, if $K<0$, making $u=\pi-2\zeta$ in the motion equation (\ref{motion}), we obtain, in similar form, that $\zeta=\pi/2$ is stable equilibria and $\zeta=0$ is unstable equilibria.

Moreover, it is well known that the equation (\ref{planar}) can be exactly solved in term of a elliptic integral of first kind (see \cite{San}, for instance), which cannot be evaluated in a closed form.

A first approximation to this problem in arbitrary constant curvature is given for small oscillations around the stable equilibria points of this physical system. That is, for small $u$, $\sin_K u 
\simeq u$, and the equation of motion is approximated by

$$\frac{d^2u}{dt^2}=-v^2 \mid K \mid u.$$ Observe that the previous equation represents a simple harmonic oscillator of frequence $\omega = v \sqrt{\mid K \mid}$.
}
\end{rema}

\subsection{A first integral}
Since the Hamiltonian function ${\cal H}$ is time independent, it is an integral on  phase space and it represents
 the system energy. Using the standard notation $(q_\zeta,p_\zeta)$ in the phase space, we can write

$$p_\zeta=\frac{\partial {\cal L}}{\partial\dot\zeta}=m\dot\zeta \sin^2_K(\rho) \frac{1}{\mid K\mid}.$$

The Legendre transformations allow us to give

\begin{equation} \label{hamiltonian}
{\cal H}(\zeta ,\dot\zeta)=\frac{1}{2}m\dot{\zeta}^2\sin_K^2(\rho)\frac{1}{\mid K\mid}+\frac{1}{2}mv^2{\rm sgn}(K)\sin_K^2(\rho)\sin^2(\zeta).
\end{equation}

Moreover, it is easy to see that ${\cal H}$ is constant if and only if

$$2\Big(\frac{d\zeta}{dt}\Big)^2-K v^2\cos(2\zeta)$$

\noindent is constant (compare with \cite[Prop. 1]{C-F-G}).

\section{Accelerated pivot}
Suppose now that the pivot moves along a geodesic path with lineal acceleration $a(t)$. We come to compute the acceleration induced by the accelerated pivot at the mass making use of a geometric argument. 
We follow the following steps:

\vspace{1mm}

\noindent a) Translate to the mass the pivot acceleration $a(t)$ .

\noindent b) Transform this acceleration in angular acceleration between the rod and the geodesic path traced by the pivot.

\noindent c) Project on the orthogonal direction to the rod.

\

\begin{figure}
\centering
\includegraphics[width=0.47\linewidth]{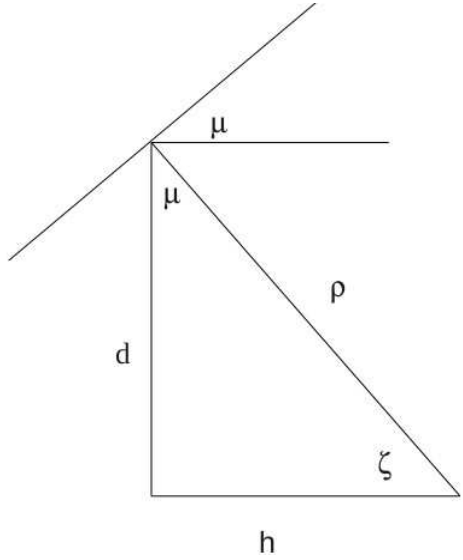}
\caption{Triangle}
\label{fig 4}
\end{figure}

\noindent Indeed, consider the geodesic triangle of the figure 4. The translated acceleration to the mass is given by
$$a(t)\cos_K(d),$$ 

\noindent and the angular acceleration by
$$\frac{a(t)\cos_K(d)}{R\sin_K(\rho)}.$$

\noindent Finally, in order to project, it is enough to multiply by $\cos(\mu)$
$$\frac{a(t)\cos_K(d)\cos(\mu)}{R\sin_K(\rho)}.$$

On another hand, taking into account the previous triangle, from the Pythagoras and Cosine Theorem in non-Euclidean geometries, we have
$$\cos_K(\mu)=\sin_K(\zeta)\cos_K(h)\ \ {\rm and}\ \ \cos_K(\rho)=\cos_K(d)\cos_K(h),$$

\noindent thus $\cos_K(d)\cos_K(\mu)=\cos_K(\rho)\sin_K(\zeta)$. As a consequence, the searched acceleration is

$$\sqrt{\mid K\mid} a(t)\cot_K(\rho)\sin_K(\zeta(t)).$$

So, we prove in a different approach the result (\cite[Theorem D]{C-F-G})

\begin{teor} Assume that the pivot moves with speed $v(t)$ along a geodesic on a surface with constant curvature $K$. Let $\zeta(t)$ the angle that the rigid rod makes with the direction of the motion of the pivot. 
Then the pendulum satisfies the differential equation
$$\frac{d^2\zeta}{dt^2}=-v(t)^2 K \sin(\zeta(t)) \cos(\zeta(t))+\sqrt{\mid K\mid} \, a(t) \, {\rm cotan}_K(\rho) \, \sin(\zeta(t)),$$

\noindent where $a(t)=\frac{dv}{dt}.$
\end{teor}

\section{Some further physical applications}

As a first application,
consider us the case of a pendulum whose pivot moves along a geodesic on a surface of constant curvature with constant speed $v$ and whose rod is elastic. Let $k$ be the  elastic constant of the rod, 
let $\rho$  be the length of the rod and let $l(t)$ be the elongation of the rod at the time $t$. Our analytical approach allows us to initiate the study of dynamical system in similar form as the rigid case. So, it is not difficult to see that
its  Lagrangian function is given by

$$
{\cal L} (\zeta,l,\dot{\zeta},\dot{l}) = \frac{1}{2} m \dot{\zeta} \sin_K^2 (\rho + l) \frac{1}{|K|} -\frac{1}{2} m v^2 {\rm sgn}(K)  \sin^2_K (\rho + l) \sin_K^2 (\zeta) + \frac{1}{2} m \dot{l}^2 - \frac{1}{2} k l^2
$$

Secondly, we will make an approximation to the quantum system.
Denote $\overline{m}:=m \sin_K^2 (\rho )/\mid K \mid$. Then, (\ref{hamiltonian}) can be written, for $K >0$, in terms of $\zeta$ and $p_\zeta$ as
$$
{\cal H}(\zeta, p_\zeta)=\frac{p_\zeta^2}{2 \overline{m} }+\frac{1}{2} \overline{m} v^2 K \sin^2(\zeta).
$$ Observe that the same expression, up an additive term, is obtained for $K<0$.
Therefore, a suitable time independient Schrodinger equation for this system can be described as

$$
- \frac{ \hbar^2 }{2 \overline{m} } \frac{d^2 \psi}{d\zeta^2} + \frac{1}{2} \overline{m} v^2 K \sin^2(\zeta) \psi = E \psi,
$$ where $\psi(\zeta)$ is a complex function representing the wave function of the mass particle in this physical system. Observe that it is
expected that the energy should take quantized values. In fact, if it is assumed that $\psi$ is confined in a region where $\zeta$ is close to $0$, the
previous equation, in this first approximation, corresponds to the quantum harmonic oscillator, whose energy is found to be
$$
E_n = \hbar v \sqrt{|K|} \left(n+\frac{1}{2} \right) , \ \ {\rm for \ \ all} \ \ n\in\mathbb{N}.
$$ Observe that, within this simplification, in the quantum of energy, $\hbar  v \sqrt{|K|}$, appears the speed of the rod and the
curvature of the ambient space.

\section*{Acknowledgments}
The authors are partially supported by the Spanish MEC-FEDER Grant MTM2010-18099.

\end{document}